\documentclass[journal]{IEEEtran}
\usepackage{graphicx}
\usepackage{balance}
\usepackage{cite}
\usepackage{multicol}
\usepackage{amsmath,amssymb,amsfonts}
\usepackage{algorithmic}
\usepackage{textcomp}
\usepackage{caption}

\newtheorem{thm}{\hskip 0.00em Theorem}

\newtheorem{lem}{\hskip 0.00em Lemma}

\newtheorem{df}{\hskip 0.00em Definition}

\newtheorem{exm}{\hskip 0.00em Example}

\ifCLASSINFOpdf
\else
\fi

\begin{document}

\title{Limited-budget output consensus for descriptor multiagent systems with energy constraints}

\author{Jianxiang~Xi,~Cheng~Wang,~Xiaojun~Yang,~Bailong~Yang\vspace{-1.5em}
\thanks{Jianxiang~Xi,~Cheng~Wang,~Bailong~Yang are with Rocket Force University of Engineering, Xi'an, 710025, P.R. China, and Xiaojun~Yang is with School of Information Engineering, Guangdong University of Technology, Guangzhou, 510006, P.R. China. (Corresponding author: Xiaojun~Yang, yangxj18@gdut.edu.cn)}}

\markboth{IEEE Transactions on Cybernetics}
{Xi \MakeLowercase{\textit{et al.}}: Limited-budget output consensus for descriptor multi-agent systems with energy constraints}

\maketitle

\begin{abstract}
The current paper deals with limited-budget output consensus for descriptor multiagent systems with two types of switching communication topologies; that is, switching connected ones and jointly connected ones. Firstly, a singular dynamic output feedback control protocol with switching communication topologies is proposed on the basis of the observable decomposition, where an energy constraint is involved and protocol states of neighboring agents are utilized to derive a new two-step design approach of gain matrices. Then, limited-budget output consensus problems are transformed into asymptotic stability ones and a valid candidate of the output consensus function is determined. Furthermore, sufficient conditions for limited-budget output consensus design for two types of switching communication topologies are proposed, respectively. Finally, two numerical simulations are shown to demonstrate theoretical conclusions.\end{abstract}


\IEEEpeerreviewmaketitle

\section{Introduction}\label{sec:introduction}
\IEEEPARstart{D}{uring} the last decade, consensus of multiagent systems receives considerable attention, which designs a distributed control protocol to drive multiple agents to achieve an agreement about some interested variables such as time, position, velocity and temperature,  {\it et al}., as shown in \cite{b1}-\hspace{-0.5pt}\cite{b6}. Consensus has potential practical applications in formation control \cite{b7}-\hspace{-0.5pt}\cite{b10}, target tracking \cite{b11}-\hspace{-0.5pt}\cite{b12}, network synchronisation \cite{b13}-\hspace{-0.5pt}\cite{b14} and multiple source data analysis \cite{b15}-\hspace{-0.5pt}\cite{b16}, {\it et al}.\par

The communication topologies are critically important for multiagent systems to achieve consensus, which can usually be categorized into the fixed ones and switching ones. For fixed communication topologies, the neighboring relationship and communication weights are time-invariant, as discussed in \cite{b17} and \hspace{-0.5pt}\cite{b18}. For switching communication topologies, the neighboring relationships may be time-varying, but communication weights are time-invariant. In this case, the associated Laplacian matrices of the communication topologies are piecewise continuous, as shown in \cite{b19}-\hspace{-0.5pt}\cite{b21}, where switching communication topologies are divided into switching connected ones and jointly connected ones. For switching connected communication topologies, each topology in the switching set is connected. For jointly connected communication topologies, the union of topologies in certain time interval is connected, but each topology in the switching set can be unconnected. It is well-known that consensus for jointly connected communication topology cases is more complex than the one for switching connected communication topology cases.\par

According to the dynamics of each agent, multiagent systems can be categorized into normal ones and descriptor ones. For normal multiagent systems, each agent is modeled as differential equations and only owns dynamic modes, as addressed in \cite{b22}-\hspace{-0.5pt}\cite{b24}. For descriptor multiagent systems, the dynamics of each agent may contain algebraic constrains and cannot be modeled by differential equations. In practical applications, descriptor multiagent systems can be used in complex networks and multiagent supporting systems, as discussed in \cite{b25}. Moreover, the motion of each agent may be derived by impulsive modes, static modes and dynamic modes, where impulsive modes should be eliminated since they destroy the operation of the whole system, and the motions associated with both static modes and dynamic modes should achieve consensus. Yang and Liu \cite{b26} proposed necessary and sufficient consensus conditions of descriptor multiagent systems with fixed communication topologies by a static output feedback control protocol, where a specific rank constrain is required. In \cite{b27}, the impacts of communication delays on consensus of descriptor multiagent systems were analyzed, where it was also supposed that communication topologies are fixed. Wang and Huang \cite{b28} dealt with consensus analysis and design problems for descriptor multiagent systems with switching communication topologies.\par

Based on different interested variables, consensus can be categorized into state consensus and output consensus, where it is only required that partial states or state combinations achieve agreement for output consensus. Chopra and Spong \cite{b29} first introduced the concept of output consensus and proposed the associated consensus criteria for normal nonlinear multiagent systems with relative degree one. Xiao {\it et al}. \cite{b30} modeled the dynamics of each agent in multiagent systems as a normal second-order integrator and gave sufficient conditions for output consensus. Liu and Jia \cite{b31} proposed an $H_{\infty}$ analysis approach to investigate output consensus for normal multiagent systems with external disturbances, where each agent was modeled as a type of special controllability canonical form. In \cite{b32}, output consensus for normal single-input-single-output high-order linear multiagent systems was addressed by the output regulation theory. Xi  {\it et al}. \cite{b33} focused general normal high-order multiagent systems and proposed output consensus analysis and design criteria by a partial stability method. It should be pointed that the dynamics of each agent is normal in \cite{b29}-\hspace{-0.5pt}\cite{b33}, where it was shown that output consensus is more challenging than state consensus.\par

In the above literatures, the energy limitation and/or the consensus performance were not considered, which can be modeled as certain optimization consensus problems. Cao and Ren \cite{b34} gave optimal consensus criteria by the linear quadratic regulator, where it was required that the communication topology is complete, which means that the nonzero eigenvalues of the corresponding Laplacian matrix are identical. Actually, this requirement of optimal consensus on the communication topology is conservatism, and suboptimal consensus was extensively studied. Guaranteed-performance consensus for normal multiagent systems was dealt with in \cite{b35} and \cite{b36}, where it was supposed that communication topologies are fixed. Linear matrix inequality criteria for guaranteed-cost consensus with communication delays were presented in \cite{b37} and \cite{b38}. In \cite{b39} and \cite{b40}, guaranteed-cost consensus for normal multiagent systems with switching topologies was discussed and the associated guaranteed-cost consensus analysis and design criteria were proposed. In \cite{b35}-\hspace{-0.5pt}\cite{b40}, state consensus instead of output consensus for normal multiagent systems was investigated and the guaranteed cost cannot be given previously. To the best of our knowledge, limited-budget output consensus for descriptor multiagent systems with switching communication topologies is still open and the following three challenging problems should be dealt with: (i) How to eliminate impulsive modes and guarantee that the motions associated with static modes achieve output consensus; (ii) How to introduce the given budget into consensus criteria and construct the relationship between the given budget and initial outputs instead of initial states; (iii) How to determine the impacts of switching topologies and ensure the checkable feature of the associated consensus criteria.\par

The current paper focuses on limited-budget output consensus for descriptor multiagent systems and addresses the impacts of two types of switching communication topologies. Based on the observable decomposition of the dynamics of each agent, a singular dynamic output feedback consensus protocol with an energy constraint and topology switching is presented, which can make the dynamics of the whole multiagent system satisfy some separation feature; that is, its gain matrices lies on diagonal boxes of an upper triangular matrix. Furthermore, using the characteristics that the row sum of the Laplaican matrix of the communication topology is zero, the disagreement and consensus dynamics of a descriptor multiagent system are determined and they are independent with each other. For switching connected communication topology cases, a new two-step design approach is proposed to design gain matrices of consensus protocols, the relationship between the given energy budget and the matrix variable is constructed, and limited-budget output consensus design criteria are given, respectively. For jointly connected communication topology cases, by the Cauchy convergence criterion and Barbalat's lemma, sufficient conditions for limited-budget output consensus design are presented, respectively. \par

\section{Problem description}\label{sec:2}
The dynamics of the $m$th agent of a high-order homogeneous descriptor multiagent system is modeled by
\begin{equation}\label{eq:1}
\left\{\begin{array}{l}{E \dot{x}_{m}(t)=A x_{m}(t)+B u_{m}(t),} \\ {y_{m}(t)=C x_{m}(t),}\end{array}\right.
\end{equation}
where  $m \hspace{-0.5pt} =\hspace{-0.5pt} 1,2,\cdots,N, \quad E\in\mathbb{R}^{n\times n}, \quad  A\in\mathbb{R}^{n \times n}, \quad B\in\mathbb{R}^{n \times k}, \quad C \in \mathbb{R}^{l \times n}$  and $x_m(t)$, $u_m(t)$ and $y_m(t)$  denote the system state, the control input and the system output, respectively. Because the matrix $E$  may be singular; that is, rank$(E)\leq n$ , multiagent system (1) is called the descriptor multiagent system. Compared with normal multiagent systems modeled by differential equations, multiagent system (1) may contain static modes and impulsive modes besides dynamic modes. Impulsive modes must be eliminated by designing proper consensus protocols since they destroy the operation of the whole system. For multiagent system (1), all the states associated with both dynamic modes and static modes should achieve consensus.\par
This autonomous system of multiagent system (1) can also be represented by the pair $(E,A)$, whose some specific properties compared with normal systems are listed as follows.
\begin{df}\cite{b42}\label{Definition 1}
 The pair $(E,A)$ is {\it regular} if $|\sigma E-A| \neq 0$ for some $\sigma \in \mathbb{C}$ and {\it impulse-free} if deg$(|\sigma E-A|)=\text{rank}(E)$ for $\forall \sigma \in \mathbb{C}$.
\end{df}
\begin{lem}\cite{b42}\label{Lemma 1}
The pair $(E,A)$ is regular and impulse-free if and only if $\text{rank}\left[ \begin{matrix}
   E & 0  \\
   A & E  \\
\end{matrix} \right]=n+\text{rank}(E).$\end{lem}
\begin{lem} \cite{b42}\label{Lemma 2}
The pair $(E,A)$ is regular, impulse-free and asymptotically stable if and only if there exists a matrix $R$ such that ${{E}^{T}}R={{R}^{T}}E\ge 0$ and ${{A}^{T}}R+{{R}^{T}}A<0.$
\end{lem}\par

We first introduce the observable decomposition of the tripe $(E,A,C)$. Let $U_o$  be an invertible matrix such that
$$
U_{o}^{-1} E U_{o}=\left[ \begin{array}{cc}{E_{o}} & {\mathbf{0}} \\ {E_{\tilde{o}}} & {E_{\overline{o}}}\end{array}\right], \quad U_{o}^{-1} B=\left[ \begin{array}{c}{B_{o}} \\ {B_{\overline{o}}}\end{array}\right],
$$
$$
\hspace{-12pt} U_{o}^{-1} A U_{o}=\left[ \begin{array}{ll}{A_{o}} & {\mathbf{0}} \\ {A_{\tilde{o}}} & {A_{\overline{o}}}\end{array}\right], \quad C U_{o}=\left[C_{o}, \mathbf{0}\right],
$$
where $E_{o} \in \mathbb{R}^{h \times h}, \quad A_{o} \in \mathbb{R}^{h \times h}, \quad B_{o} \in \mathbb{R}^{h \times k}, \quad C_{o} \in \mathbb{R}^{l \times h}$  and the triple  $(E_o,A_o,C_o)$ is observable. Then, the following consensus protocol with switching topologies and an energy constraint is proposed:
\begin{eqnarray}\label{eq:2}
\left\{\begin{array}{l}
 \hspace{-6pt}E_o{{{\dot{z}_m}}}(t)=\left(A_o+B_o{{K}_{u}} \right){{z_m}}(t)\hfill \\
 \hspace{-6pt} \qquad\qquad \;\text{             }-{{K}_{z}}C_o\sum\limits_{j\in {{N}_{m,\kappa (t)}}}{{{w}_{mj,\kappa (t)}}\left( {{z_j}}(t)-{{z_m}}(t) \right)} \hfill \\
 \hspace{-6pt} \qquad\qquad \;\text{             }+{{K}_{z}}\sum\limits_{j\in {{N}_{m,\kappa (t)}}}{{{w}_{mj,\kappa (t)}}\left( {{y}_{j}}(t)-{{y}_{m}}(t) \right),} \hfill \\
 \hspace{-6pt}{{u}_{m}}(t)={{K}_{u}}{z_m}(t), \\
 \hspace{-6pt}{{J}_{\text{e}}}=\sum\limits_{m=1}^{N}{\int_{0}^{+\infty }{u_{m}^{T}(t)M{{u}_{m}}(t)\text{d}t}}, \\
\end{array} \right. \hspace{-6pt}
\end{eqnarray}
where $m=1,2, \cdots, N, \;\;K_{u} \in \mathbb{R}^{k \times h}, \;\;K_{z} \in \mathbb{R}^{h \times l}, \;\;M^{T}=M>0, N_{m, \kappa(t)}$  is the neighbor set of the $m$th agent at time $t$, $z_m(t)$  with $z_m(0)=\mathbf{0}$  stands for the protocol state, and  $J_{\text{e}}$ represents the energy consumption of the whole multiagent system.\par

Let $J_{\text{e}}^{*}>0$  be a given energy budget; that is, the whole energy consumption of multiagent system (1) with consensus protocol (2) must be less than  $J_{\text{e}}^{*}$. In the following, we give the definition of the limited-budget output consensus of descriptor multiagent systems.\par

\begin{df}\label{Definition 2}
 For any given $J_{\text{e}}^{*}>0$, multiagent system (1) is said to be {\it limited-budget output consensualizable} by consensus protocol (2) if it is regular and impulse-free and there exist $K_u$  and  $K_z$ such that  $\lim _{t \rightarrow+\infty}\left(y_{m}(t)-c_{o}(t)\right)=\mathbf{0} \hspace{3pt}(m=1,2,\cdots ,N)$   and $J_{\text{e}} \leq J_{\text{e}}^{*}$  for any bounded disagreement initial outputs, where $c_o(t)$ is said to be {\it the output consensus function}.
\end{df}

The current paper gives two design approaches of gain matrices  $K_u$ and $K_z$  such that multiagent system (1) with consensus protocol (2) achieves limited-budget output consensus for switching connected communication topologies and jointly connected communication topologies, respectively. 

\section{Problem transformation}\label{section3}
By the observable decomposition and the separation principle, consensus problems for multiagent system (1) with consensus protocol (2) are transformed into asymptotic stability ones for a reduced-order subsystem, and the consensus dynamics is also determined.\par

Let  $U_{o}^{-1} x_{m}(t)=\left[x_{o m}^{T}(t), x_{\overline{o} m}^{T}(t)\right]^{T}$, then multiagent system (1) can be transformed into
\begin{eqnarray}\label{eq:3}
\left\{\begin{array}{l}
 \vspace{2pt}
 \hspace{-0.15cm} \left[ \begin{array}{cc}{E_{o}} & {\mathbf{0}} \\ {E_{\tilde{o}}} & {E_{\overline{o}}}\end{array}\right]
\hspace{-4pt} \left[ \begin{array}{cc}{\dot{x}_{om}(t)} \\ {\dot{x}_{\overline{o}m}(t)} \end{array}\right]
=\left[ \begin{array}{ll}{A_{o}} & {\mathbf{0}} \\ {A_{\tilde{o}}} & {A_{\overline{o}}}\end{array}\right]
\hspace{-4pt}\left[ \begin{array}{cc}{{x}_{om}(t)} \\ {{x}_{\overline{o}m}(t)} \end{array}\right]  \\
  \vspace{2pt}
 \hspace{3.9cm} +\left[ \begin{array}{cc}{{B}_{o}(t)} \\ {{B}_{\overline{o}}(t)} \end{array}\right]u_m(t),\\
\hspace{-0.15cm} y_m(t)=C_ox_{om}(t),
\end{array} \right.
\end{eqnarray}
where  $m=1,2,\cdots,N$. It can be found that the system output  $y_m(t) (m\in{1,2,\cdots,N})$   depends on the observable component  $x_{om}(t)$, but it is not associated with the unobservable component   $x_{\overline{o}m}(t)$. Hence, when the limited budget is not considered, multiagent system (1) achieves output consensus if and only if the observable component of each agent achieves consensus. In this case, the unobservable component of each agent can be neglected and the observable component of multiagent system (1) with consensus protocol (2) can be written in the following Kronecker form as
$$
\hspace{-1.3cm}I_{N} \otimes \left[\begin{array}{cc}{E_{o}} & { } \\ { } & {E_{o}}\end{array}\right]\hspace{-0.1cm} \left[ \begin{array}{c}{\dot{x}_{o}(t)} \\ {\dot{z}_{o}(t)}\end{array}\right]=
\left[ \begin{array}{c}{I_{N} \otimes A_{o}} \\ {-L_{\kappa(t)} \otimes K_{z} C_{o}}\end{array}\right.
$$
\begin{equation}\label{eq:4}
\hspace{0.45cm}\left.\begin{array}{c}{I_{N} \otimes B_{o} K_{u}} \\ {I_{N} \otimes\left(A_{o}+B_{o} K_{u}\right)+L_{\kappa(t)} \otimes K_{z} C_{o}}\end{array}\right]
\hspace{0pt} \left[ \begin{array}{l}{x_{o}(t)} \\ {z_{o}(t)}\end{array}\right],
\end{equation}
where  $x_{o}(t)=\left[x_{o 1}^{T}(t), x_{o 2}^{T}(t), \cdots, x_{o N}^{T}(t)\right]^{T}, \quad z_{o}(t)=\left[z_{1}^{T}(t), z_{2}^{T}(t), \cdots, z_{N}^{T}(t)\right]^{T}$  and $L_{\kappa}(t)$  is the Laplacian matrix of the switching communication topology, which is piecewise continuous since the switching signal $\kappa(t)$  is piecewise continuous.\par

Since each communication topology in the switching set is undirected, the associated Laplacian matrix  $L_{\kappa(t)}$ is symmetric, where zero is its eigenvalue with an eigenvector $\mathbf{1}/\sqrt{N}$. In this case, there exists an orthonormal matrix  $U_{\kappa}=\left[\mathbf{1} / \sqrt{N}, \tilde{U}_{\kappa}\right]$ such that
 $$
U_{\kappa}^{T} L_{\kappa(t)} U_{\kappa}=\operatorname{diag}\left\{0, \Delta_{\kappa(t)}\right\},
$$
where  $\Delta_{\kappa(t)}=\tilde{U}_{\kappa}^{T} L_{\kappa(t)} \tilde{U}_{\kappa} \in \mathbb{R}^{(N-1) \times(N-1)}$ and $U_{\kappa}$  is set to be time-varying for switching connected communication topology cases and time-invariant for jointly connected communication topology cases. We introduce the following nonsingular transformation
$$
\tilde{x}_{o}(t)=\left(U_{\kappa}^{T} \otimes I_{h}\right) x_{o}(t)=\left[\tilde{x}_{o 1}^{T}(t), \tilde{x}_{o 2}^{T}(t), \cdots, \tilde{x}_{o N}^{T}(t)\right]^{T},
$$
$$
\tilde{z}_{o}(t)=\left(U_{\kappa}^{T} \otimes I_{h}\right) z_{o}(t)=\left[\tilde{z}_{o 1}^{T}(t), \tilde{z}_{o 2}^{T}(t), \cdots, \tilde{z}_{o N}^{T}(t)\right]^{T}.
$$\par

\hspace{-7pt}Thus, multiagent system (4) can be transformed into
 \begin{equation}\label{eq:5}
\left[\hspace{-2pt}\begin{array}{cc}{E_{o}}\hspace{-7pt}&{ } \\ { }\hspace{-7pt}&{E_{o}}\end{array}\hspace{-2pt}\right]\hspace{-4pt}  \left[ \begin{array}{c}{\dot{\tilde{z}}_{o 1}(t)} \\ {\dot{\tilde{x}}_{o 1}(t)}\end{array}\hspace{-2pt}\right] \hspace{-3pt}= \hspace{-3pt}\left[\hspace{-2pt}\begin{array}{cc}{A_{o}+B_{o} K_{u}} \hspace{-7pt}& {\mathbf{0}} \\ {B_{o} K_{u}}\hspace{-7pt} & {A_{o}}\end{array}\hspace{-2pt}\right]\hspace{-4pt} \left[ \begin{array}{c}{\tilde{z}_{o 1}(t)} \\ {\tilde{x}_{o 1}(t)}\end{array}\right]\hspace{-2pt},\hspace{-2pt}
\end{equation}
\vspace{-0.1cm}
$$
\hspace{-4.1cm}I_{N-1}\hspace{-2pt}\otimes\hspace{-2pt}\left[\hspace{-2pt}\begin{array}{cc}{E_{o}} & { } \\ { } & {E_{o}}\end{array}\hspace{-2pt}\right]
\left[\hspace{-2pt}\begin{array}{c}{\dot{\tilde{z}}_{o \Delta}(t)} \\ {\dot{\tilde{x}}_{o \Delta}(t)}\end{array}\right]
$$
\vspace{-0.2cm}
$$
=\left[\hspace{-4pt}\begin{array}{cc}{I_{N-1}\hspace{-2pt}\otimes\hspace{-2pt}\left(A_{o}+B_{o} K_{u}\right)
\hspace{-2pt}+\hspace{-2pt}\Delta_{\kappa(t)}\hspace{-2pt}\otimes\hspace{-2pt}K_{z} C_{o}}\hspace{-4pt}& {-\Delta_{\kappa(t)}
\otimes K_{z} C_{o}} \\ {I_{N-1}\hspace{-2pt}\otimes\hspace{-2pt}B_{o} K_{u}}\hspace{-4pt}& {I_{N-1}\hspace{-2pt}\otimes\hspace{-2pt}A_{o}}\end{array}\hspace{-4pt}\right]
$$
\vspace{-0.2cm}
\begin{equation}\label{eq:6}
\hspace{-5.8cm}\times\left[ \begin{array}{l}{\tilde{z}_{o \Delta}(t)} \\ {\tilde{x}_{o \Delta}(t)}\end{array}\right],
\end{equation}
where $\tilde{x}_{o \Delta}(t)\hspace{3pt}=\hspace{3pt}\left[\begin{array}{l}{\tilde{x}_{o 2}^{T}(t),\hspace{2pt}\tilde{x}_{o 3}^{T}(t),\hspace{2pt}\cdots,\hspace{2pt}\tilde{x}_{o N}^{T}(t)\hspace{4pt}]^{T}}\end{array}\right.$ and $\quad \tilde{z}_{o \Delta}(t)\hspace{-1pt}=\hspace{-1pt}\left[\begin{array}{c}{\tilde{z}_{o 2}^{T}(t),\hspace{2pt}\tilde{z}_{o 3}^{T}(t),\hspace{2pt}\cdots,\hspace{2pt} \tilde{z}_{o N}^{T}(t)\hspace{4pt}]^{T}}\end{array}\right.$\hspace{-6pt}.  Let
 $$
U_{s}=\left[ \begin{array}{cc}{I_{(N-1) h}} & {\mathbf{0}} \\ {I_{(N-1) h}} & {-I_{(N-1) h}}\end{array}\right],
$$
$$
\left[ \begin{array}{c}{\hat{z}_{o \Delta}(t)} \\ {\hat{x}_{o \Delta}(t)}\end{array}\right]=U_{s} \left[ \begin{array}{c}{\tilde{z}_{o \Delta}(t)} \\ {\tilde{x}_{o \Delta}(t)}\end{array}\right],
$$
then subsystem (6) can be transformed into
$$
\hspace{-4.1cm}I_{N-1}\hspace{-2pt}\otimes\hspace{-2pt}\left[\hspace{-2pt}\begin{array}{cc}{E_{o}} & { } \\ { } & {E_{o}}\end{array}\hspace{-2pt}\right]
\left[ \begin{array}{l}{\dot{\hat{z}}_{o \Delta}(t)} \\ {\dot{\hat{x}}_{o \Delta}(t)}\end{array}\right]
$$
\vspace{-0.2cm}
$$
=\left[\hspace{-4pt}\begin{array}{cc}{I_{N-1}\hspace{-2pt}\otimes\hspace{-2pt}\left(A_{o}+B_{o} K_{u}\right)}
 & {\Delta_{\kappa(t)}\otimes K_{z} C_{o}} \\ {\mathbf{0}} &
{I_{N-1}\hspace{-2pt}\otimes\hspace{-2pt}A_{o}
\hspace{-2pt}+\hspace{-2pt}\Delta_{\kappa(t)}\hspace{-2pt}\otimes\hspace{-2pt}K_{z} C_{o}}\hspace{-4pt}\end{array}\hspace{-1pt}\right]
$$
\vspace{-0.2cm}
\begin{equation}\label{eq:7}
\hspace{-5.0cm}\times\left[ \begin{array}{l}{\hat{z}_{o \Delta}(t)} \\ {\hat{x}_{o \Delta}(t)}\end{array}\right].
\end{equation}\par

Let the minimum and maximum nonzero eigenvalues of the Laplacian matrices of all the communication topologies in the switching set denote as $\lambda_{\text{min}}=\text{min}\{\lambda_{m,\text{min}},\forall m\in\chi\}$  and $\lambda_{\text{max}}=\text{max}\{\lambda_{m,\text{min}},\forall m\in\chi\},$   where $\lambda_{m,\text{min}}$  and   $\lambda_{m,\text{max}}$ represent the minimum and maximum nonzero eigenvalues of the Laplacian matrix of the $m$th communication topology in the switching set, respectively. These two notations are used for both switching connected communication topology cases and jointly connected communication topology cases.\par

Moreover, subsystems (5) and (6) determine consensus and disagreement dynamics of multiagent system (1). The following theorem converts the consensus problem into the asymptotic stability one and determines a candidate of the output consensus function.\par

\begin{thm}\label{Theorem 1}
 If multiagent system (1) with consensus protocol (2) is regular and impulse-free and  $\lim _{t \rightarrow+\infty}\left[\hat{z}_{o \Delta}^{T}(t),\right.$ $\hat{x}_{o \Delta}^{T}(t)\left]^T\right.=\mathbf{0}$, then it achieves output consensus and $\lim _{t \rightarrow \infty}\left(c_{o}(t)-C_{o} \tilde{x}_{o 1}(t) / \sqrt{N}\right)=\mathbf{0}$.
\end{thm}\par

{\it Proof}: Let $e_m (m\in{1,2,\cdots,N})$ denote the $N$-dimensional column vector with the $m$th component 1 and 0 elsewhere. Due to
$$
U_{\kappa}e_{1}=\frac{1}{\sqrt{N}}\mathbf{1},
$$
one can show that
\begin{equation}\label{eq:8}
U_{\kappa} e_{1} \otimes \tilde{x}_{o 1}(t)=\frac{1}{\sqrt{N}} \mathbf{1} \otimes \tilde{x}_{o 1}(t).
\end{equation}
From the property of the Kronecker product, one can find that
  \begin{equation}\label{eq:9}
U_{\kappa} e_{1} \otimes \tilde{x}_{o 1}(t)=\left(U_{\kappa} \otimes I_{h}\right) \left[ \begin{array}{c}{\tilde{x}_{o 1}(t)} \\ {\mathbf{0}}\end{array}\right],
\end{equation}
\begin{equation}\label{eq:10}
\sum_{m=2}^{N} U_{\kappa} e_{m} \otimes \tilde{x}_{o m}(t)=\left(U_{\kappa} \otimes I_{h}\right) \left[ \begin{array}{c}{\mathbf{0}} \\ {\tilde{x}_{o \Delta}(t)}\end{array}\right].
\end{equation}
Because $U_s$  is invertible, $\lim _{t \rightarrow \infty}\left[\tilde{z}_{o \Delta}^{T}(t), \tilde{x}_{o \Delta}^{T}(t)\right]^{T}=\mathbf{0}$  if  $\lim _{t \rightarrow \infty}\left[\hat{z}_{o \Delta}^{T}(t), \hat{x}_{o \Delta}^{T}(t)\right]^{T}=\mathbf{0}$. Due to
$$
x_{o}(t)=\left(U_{\kappa} \otimes I_{h}\right) \tilde{x}_{o}(t),
$$
the conclusion can be obtained from (8) to (10). The proof of Theorem 1 is completed.$\blacksquare$
\section{Switching connected communication topology cases}\label{section4}
For multiagent system (1) with switching connected communication topologies, this section gives sufficient conditions for limited-budget output consensualization and consensus in terms of matrix inequality techniques.\par

Since each communication topology in the switching set is connected, zero is its simple eigenvalue of the Laplacian matrix $L_{\kappa}(t)$  and all other eigenvalues are positive. Without loss of generality, we can set that $\tilde{U}_{\kappa}=\tilde{U}_{\kappa(t)}$  such that $\tilde{U}_{\kappa(t)}^{T} \Delta_{\kappa(t)} \tilde{U}_{\kappa(t)}=\operatorname{diag}\left\{\lambda_{\kappa(t), 2}, \lambda_{\kappa(t), 3}, \cdots, \lambda_{\kappa(t), N}\right\}$  with  $0<\lambda_{\kappa(t), 2} \leq \lambda_{\kappa(t), 3} \leq \cdots \leq \lambda_{\kappa(t), N}$. In this case,  $\tilde{U}_{\kappa}$ is piecewise continuous, $\lambda_{\text{min}}=\text{min}\{\lambda_{m,2},\forall m\in\chi\}$  and $\lambda_{\text{max}}=\text{max}\{\lambda_{m,N},\forall m\in\chi\}$.  Let  $y_{o}(0)=\left[y_{1}^{T}(0), y_{2}^{T}(0), \cdots, y_{N}^{T}(0)\right]^{T}$. The following theorem proposes a limited-budget output consensualization criterion by matrix inequality techniques.\par

\begin{thm}\label{Theorem 2}
 For any given $J_\text{e}^{*}>0$, multiagent system (1) with switching connected cummnication topologies is limited-budget output consensualizable by consensus protocol (2) if there exist $R_x$  and  $R_z$ such that
 $$
\hspace{-3.2cm}(\mathrm{I})\text{rank}\left[ \begin{array}{ll}{E_{o}} & {\mathbf{0}} \\ {A_{o}} & {E_{o}}\end{array}\right]=h+\text{rank}(E_o),
$$
$$
(\mathrm{II}) \left\{\begin{array}{l}{E_{o}^{T} R_{x}=R_{x}^{T} E_{o} \geq {0}}, \\ {R_{x}^{T} A_{o}+A_{o}^{T} R_{x}-C_{o}^{T} C_{o}<0}, \\
{y_{o}^{T}(0)\left(\left(I_{N}\hspace{-2pt}-\hspace{-2pt}N^{-1} \mathbf{1}
\mathbf{1}^{T}\right)\hspace{-2pt}\otimes\hspace{-2pt}I_{h}\right) y_{o}(0) E_{o}^{T} R_{x}
\hspace{-2pt}\leq\hspace{-2pt} J_\text{e}^{*}C_{o}^{T} C_{o}},\end{array}\right.
$$
$$
\hspace{-1.82cm}(\mathrm{III}) \left\{\begin{array}{l}{R_{z}^{T} E_{o}^{T}=E_{o} R_{z} \geq 0},\\
\Theta_m=\left[\begin{array}{ccc}{\Theta_{11}} & {\Theta_{12}^{m}} & {0.5 B_{o} M} \\ {*} & {\Theta_{22}} & {\mathbf{0}} \\ {*} & {*} & {-M}\end{array}\right]<0,
\end{array}\right.
 $$
 \end{thm}
 where $\lambda_m=\lambda_{\text{min}},\lambda_{\text{max}}$ and
 $$
\begin{aligned} \Theta_{11} &=A_{o} R_{z}+R_{z}^{T} A_{o}^{T}-B_{o} B_{o}^{T}, \\ \Theta_{12}^{m} &=-0.5 \lambda_{m} R_{x}^{-T} C_{o}^{T} C_{o}, \\ \Theta_{22} &=R_{x}^{T} A_{o}+A_{o}^{T} R_{x}-C_{o}^{T} C_{o}. \end{aligned}
$$

In this case, $K_u\hspace{-3pt}=\hspace{-3pt}-B_o^TR_z^{-1}/2$ and $K_z\hspace{-3pt}=\hspace{-3pt}-\lambda_{\text{min}}^{-1}R_x^{-T}C_o^T/2$.\par

{\it Proof}: Consider a Lyapunov function candidate as follows
$$
V_{x}(t)=\hat{x}_{o \Delta}^{T}(t)\left(I_{N-1} \otimes E_{o}^{T} R_{x}\right) \hat{x}_{o \Delta}(t),
$$
where $E_{o}^{T} R_{x}\hspace{-3pt}=\hspace{-3pt}R_{x}^{T} E_{o}\hspace{-3pt}\geq\hspace{-3pt}0$. Let $\hat{x}_{o\Delta}(t)\hspace{-3pt}=\hspace{-3pt}\left[(\hat x_{o\Delta}^2(t))^T,\hat x_{o\Delta}^3(t))^T,\right.$
$\cdots,\hat x_{o\Delta}^N(t))^T\left]^T\right.$ with $\hat{x}_{o \Delta}^{m}(t) \in \mathbb{R}^{h}\hspace{2pt}(m=2,3, \cdots, N)$, then it can be deduced by (7) that
$$
\dot{V}_x(t)=\sum_{m=2}^{N}(\hat x_{o\Delta}^m(t))^T\left(R_{x}^{T}\left(A_{o}+\lambda_{\kappa(t), m} K_{z} C_{o}\right)\right.
$$
$$
\hspace{22pt}+\left(A_{o}+\lambda_{\kappa(t), m} K_{z} C_{o}\right)^{T} R_{x}\left)\right.\hat x_{o\Delta}^m(t).
$$
Let $K_z=-\lambda_{\text{min}}^{-1}R_x^{-T}C_o^T/2$, then one can obtain that
$$
\dot V_x(t)\hspace{-3pt}=\hspace{-6pt}\sum_{m=2}^{N}\hspace{-2pt}(\hat x_{o\Delta}^m(t)\hspace{-1pt})^T\hspace{-3pt}\left(\hspace{-1pt}R_{x}^{T}A_o\hspace{-3pt}+\hspace{-3pt}A_o^TR_x\hspace{-3pt}-\hspace{-3pt}\lambda_{\kappa(t),m}\lambda_{\text{min}}^{-1}C_o^TC_o\hspace{-1pt}\right)\hspace{-1pt}\hat x_{o\Delta}^m\hspace{-2pt}(t)
$$
\begin{equation}\label{eq:11}
\leq \sum_{m=2}^{N}\left(\hat{x}_{o \Delta}^{m}(t)\right)^{T}\left(R_{x}^{T} A_{o}+A_{o}^{T} R_{x}-C_{o}^{T} C_{o}\right) \hat{x}_{o \Delta}^{m}(t).
\end{equation}
Furthermore, construct the following Lyapunov function candidate
$$
V_{z}(t)=\hat{z}_{o \Delta}^{T}(t)\left(I_{N-1} \otimes E_{o}^{T} \hat{R}_{z}\right) \hat{z}_{o \Delta}(t),
$$
where $E_{o}^{T}{{\hat{R}}_{z}}\hspace{-3pt}=\hspace{-3pt}\hat{R}_{z}^{T}{{E}_{o}}\hspace{-3pt}\ge\hspace{-3pt}0$.  Let $\hat{z}_{o \Delta}(t)\hspace{-3pt}=\hspace{-3pt}\Big[\left(\hat{z}_{o \Delta}^{2}(t)\right)^{T}\hspace{-4pt},\left(\hat{z}_{o \Delta}^{3}(t)\right)^{T}\hspace{-4pt},$ $\cdots,\left(\hat{z}_{o \Delta}^{N}(t)\right)^{T}\Big]^{T}$  with  $\hat{z}_{o\Delta }^{m}(t)\in {{\mathbb{R}}^{h}}\text{ }\left( m=2,3,\cdots ,N \right)$, then it can be deduced by (7) that
$$
\begin{aligned} \dot{V}_{z}(t)
&=\sum_{m=2}^{N}\left(\hat{z}_{o \Delta}^{m}(t)\right)^{T}\left(\hat{R}_{z}^{T}\left(A_{o}+B_{o} K_{u}\right)\right.
\\ &\hspace{10pt}+\left(A_{o}+B_{o} K_{u}\right)^{T} \hat{R}_{z} \Big)\hat{z}_{o \Delta}^{m}(t)
\\ &\hspace{10pt}+2 \lambda_{\kappa(t), m}\left(\hat{z}_{o \Delta}^{m}(t)\right)^{T} \hat{R}_{z}^{T} K_{z} C_{o} \hat{x}_{o \Delta}^{m}(t).
\end{aligned}
$$
Let $K_u=-B_o^TR_z^{-1}/2$ and  $R_z=\hat R_z^{-1}$, then it can be derived that
$$
\dot{V}_{z}(t)\hspace{-3pt}=\hspace{-5pt}\sum_{m=2}^{N}\left(R_{z}^{-1} \hat{z}_{o \Delta}^{m}(t)\right)^{T}\hspace{-3pt}\left(A_{o} R_{z}\hspace{-3pt}+\hspace{-3pt}R_{z}^{T} A_{o}^{T}\hspace{-3pt}-\hspace{-3pt}B_{o} B_{o}^{T}\right)\hspace{-3pt}\left(R_{z}^{-1} \hat{z}_{o \Delta}^{m}(t)\right)
$$
\begin{equation}\label{eq:12}
\hspace{-1.1cm}+2\lambda_{\kappa(t), m}\left(R_{z}^{-1} \hat{z}_{o \Delta}^{m}(t)\right)^{T} K_{z} C_{o} \hat{x}_{o \Delta}^{m}(t).
\end{equation}
Because $U_{\kappa(t)}\hspace{-3pt}\otimes\hspace{-3pt}I_{h}$  and  $U_s$ are nonsingular, multiagent system (4) is regular and impulse-free if and only if  $(E_o,A_o)$,   $\left(E_{o}, A_{o}+\lambda_{\kappa(t), m} K_{z} C_{o}\right) (m=2,3, \cdots, N)$ and  $(E_o,A_o+B_oK_u)$ are regular and impulse-free. From Lemma 1, the pair $(E_o,A_o)$  is regular and impulse-free if Condition (I) holds. By Lemma 2,  $\left(E_{o}, A_{o}+\lambda_{\kappa(t), m} K_{z} C_{o}\right) (m=2,3, \cdots, N)$   are regular and impulse-free if $E_{o}^{T} R_{x}=R_{x}^{T} E_{o} \geq 0$  and  $R_{x}^{T} A_{o}+A_{o}^{T} R_{x}-C_{o}^{T} C_{o}<0$, and $(E_o,A_o+B_oK_u)$  is regular and impulse-free if $R_z^TE_o^T=E_oR_z\geq 0$  and  $A_{o} R_{z}+R_{z}^{T} A_{o}^{T}-B_{o} B_{o}^{T}<0$. Thus, by (11) and (12), according to Theorem 1, multiagent system (1) with consensus protocol (2) achieves output consensus if Conditions (I)-(III) hold.\par
In the sequel, the limited-budget output consensus is addressed. It can be shown by (2) that
\begin{equation}\label{eq:13}
J_{\mathrm{e}}=\int_{0}^{+\infty} z_{o}^{T}(t)\left(I_{N} \otimes K_{u}^{T} M K_{u}\right) z_{o}(t) \mathrm{d} t.
\end{equation}
Due to $z_{m}(0)=\mathbf{0}\hspace{4pt}(m=1,2, \cdots, N)$  and  $\tilde{z}_{o}(t)=\left(U_{\kappa(t)}^{T} \otimes I_{h}\right) z_{o}(t)$, one has  $\tilde z_{o1}=\mathbf{0}$. If $R_{z}^{T} E_{o}^{T}=E_{o} R_{z} \geq 0$  and  $A_{o} R_{z}+R_{z}^{T} A_{o}^{T}-B_{o} B_{o}^{T}<0$, then  $\lim _{t \rightarrow \infty} \tilde{z}_{o 1}(t)=\mathbf{0}$. Hence, it can be deduced by (13) that
\begin{equation}\label{eq:14}
\hspace{0pt}J_{\mathrm{e}}\hspace{-3pt}=\hspace{-3pt}\sum_{m=2}^{N}\hspace{-1pt}\int_{0}^{+\infty}\hspace{-5pt}0.25\hspace{-3pt}\left(R_{z}^{-1} \tilde{z}_{o m}(t)\hspace{-1pt}\right)^{T}\hspace{-4pt}B_{o} M\hspace{-2pt}B_{o}^{T}\hspace{-3pt}\left(R_{z}^{-1} \tilde{z}_{o m}(t)\hspace{-1pt}\right)\hspace{-2pt} \mathrm{d}t.\hspace{-9pt}
\end{equation}
For any  $\hbar\geq 0$, it can be shown that
\begin{equation}\label{eq:15}
\int_{0}^{\hbar}\hspace{-3pt}\left(\dot{V}_{x}(t)\hspace{-3pt}+\hspace{-3pt}\dot{V}_{z}(t)\right)
\mathrm{d}t\hspace{-1pt}=\hspace{-1pt}V_{x}(\hbar)\hspace{-3pt}-\hspace{-3pt}V_{x}(0)\hspace{-3pt}+\hspace{-3pt}V_{z}(\hbar)\hspace{-3pt}-\hspace{-3pt}V_{z}(0).
\end{equation}
Due to  $\hat{z}_{o \Delta}(t)=\tilde{z}_{o \Delta}(t)$, from (14) and (15), it can be deduced that
$$
\begin{aligned}\hspace{-23pt}J_{\mathrm{e}}^{\hbar}
&\triangleq\int_{0}^{\hbar} z_{o}^{T}(t)\left(I_{N} \otimes K_{u}^{T} M K_{u}\right) z_{o}(t) \mathrm{d}t
\\& \hspace{-23pt}=\sum_{m=2}^{N} \int_{0}^{\hbar} 0.25\left(R_{z}^{-1} \hat{z}_{o m}(t)\right)^{T} B_{o} M B_{o}^{T}\left(R_{z}^{-1} \hat{z}_{o m}(t)\right) \mathrm{d} t
\end{aligned}
$$
\begin{equation}\label{eq:16}
\hspace{11pt}+\hspace{-3pt}\int_{0}^{\hbar}\left(\dot{V}_{x}(t)\hspace{-3pt}+\hspace{-3pt}\dot{V}_{z}(t)\right)
\mathrm{d}t\hspace{-3pt}+\hspace{-3pt}V_{x}(0)\hspace{-3pt}+\hspace{-3pt}V_{z}(0)\hspace{-3pt}-\hspace{-3pt}V_{x}(\hbar)\hspace{-3pt}-\hspace{-3pt}V_{z}(\hbar).
\end{equation}
By (11) and (12), one can find that
$$
\dot{V}_{x}(t)+\dot{V}_{z}(t)\hspace{-3pt}=\hspace{-3pt}\sum_{m=2}^{N} \left[ \begin{array}{c}{R_{z}^{-1} \hat{z}_{o \Delta}^{m}(t)} \\ {\hat{x}_{o \Delta}^{m}(t)}\end{array}\right]^{T}
\hspace{-3pt}\left[ \begin{array}{c}{A_{o} R_{z}\hspace{-3pt}+\hspace{-3pt}R_{z}^{T} A_{o}^{T}\hspace{-3pt}-\hspace{-3pt}B_{o} B_{o}^{T}} \\ {*}\end{array}\right.
$$
\begin{equation}\label{eq:17}
\left.\begin{array}{c}{-0.5 \lambda_{\kappa(t), m} \lambda_{\min }^{-1} R_{x}^{-T} C_{o}^{T} C_{o}} \\ {R_{x}^{T} A_{o}+A_{o}^{T} R_{x}-C_{o}^{T} C_{o}}\end{array}\right] \left[ \begin{array}{c}{R_{z}^{-1} \hat{z}_{o \Delta}^{m}(t)} \\ {\hat{x}_{o \Delta}^{m}(t)}\end{array}\right].
\end{equation}
From (16) and (17), based the convex property of the linear matrix inequality, if Condition (III) holds, then it can be derived as  $\hbar \rightarrow+\infty$ that
\begin{equation}\label{eq:18}
J_{\text{e}}\leq V_x(0)+V_z(0).
\end{equation}
Since it is assumed that  $z_{m}(0)=\mathbf{0}\hspace{3pt}(m=1,2, \cdots, N)$, one has  $\tilde{z}_{o}(0)=\left(U_{\kappa(0)}^{T} \otimes I_{h}\right) z_{o}(0)=\mathbf{0}$; which means that  $\tilde z_{o\Delta}(0)=\mathbf{0}$. According to the structure of  $U_s$, it can be obtained that
$$
\hat{z}_{o \Delta}(0)=\tilde{z}_{o \Delta}(0)=\mathbf{0},
$$
\begin{equation}\label{eq:19}
\hat{x}_{o \Delta}(0)=\tilde{z}_{o \Delta}(0)-\tilde{x}_{o \Delta}(0)=-\tilde{x}_{o \Delta}(0).
\end{equation}
Thus, one can see that
\begin{equation}\label{eq:20}
V_{x}(0)+V_{z}(0)=\hat{x}_{o \Delta}^{T}(0)\left(I_{N-1} \otimes E_{o}^{T} R_{x}\right) \hat{x}_{o \Delta}(0).
\end{equation}
From (18) to (20), it can be derived that
$$
J_{\mathrm{e}} \leq x_{o}^{T}(0)\left(U_{\kappa(0)} \otimes I_{h}\right) \left[ \begin{array}{c}{\mathbf{0}^{T}} \\ {I_{(N-1) h}}\end{array}\right]\left(I_{N-1} \otimes E_{o}^{T} R_{x}\right)
$$
\begin{equation}\label{eq:21}
\hspace{-2cm}\left[\mathbf{0}, I_{(N-1) h}\right]\left(U_{\kappa(0)}^{T} \otimes I_{h}\right) x_{o}(0).
\end{equation}
Due to  $U_{\kappa(0)} U_{\kappa(0)}^{T}=I_{N}$, one has
\begin{equation}\label{eq:22}
\tilde{U}_{\kappa(0)} \tilde{U}_{\kappa(0)}^{T}=I_{N}-\frac{1}{N} \mathbf{1 1}^{T}.
\end{equation}
Due to $\left[\mathbf{0}, I_{(N-1) h}\right]\left(U_{\kappa(0)}^{T} \otimes I_{h}\right)=\tilde{U}_{\kappa(0)}^{T} \otimes I_{h}$, one can find by (21) and (22) that
\begin{equation}\label{eq:23}
J_{\mathrm{e}} \leq x_{o}^{T}(0)\left(\left(I_{N}-\frac{1}{N} \mathbf{1 1}^{T}\right) \otimes E_{o}^{T} R_{x}\right) x_{o}(0).
\end{equation}
Furthermore, one can obtain that
$$
\hspace{-2.5cm}x_{o}^{T}(0)\left(\left(I_{N}-\frac{1}{N} \mathbf{1 1}^{T}\right) \otimes C_{o}^{T} C_{o}\right) x_{o}(0)
$$
\begin{equation}\label{eq:24}
\hspace{2.7cm}=\tilde{x}_{o \Delta}^{T}(0)\left(I_{N-1} \otimes C_{o}^{T} C_{o}\right) \tilde{x}_{o \Delta}(0).
\end{equation}
Due to  $x_{o}(0)=\left(U_{\kappa(0)} \otimes I_{h}\right) \tilde{x}_{o}(0)$, it can be obtained that
$$
y_{o}(0)=\left(U_{\kappa(0)} \otimes C_{o}\right) \tilde{x}_{o}(0).
$$
By (9) and (10), the disagreement output component at  $t=0$ is
\begin{equation}\label{eq:25}
y(0)\hspace{-2pt}-\hspace{-2pt}\frac{1}{\sqrt{N}} \mathbf{1}\hspace{-2pt}\otimes\hspace{-2pt}C_{o}(0)\tilde{x}_{o1}(0)\hspace{-2pt}=\hspace{-2pt}\left(U_{\kappa(0)}\hspace{-2pt}\otimes\hspace{-2pt}C_{o}\right) \left[\begin{array}{c}{\mathbf{0}} \\ {\tilde{x}_{o \Delta}(0)}\end{array}\right].
\end{equation}
Because it is supposed that $y_m(0)\hspace{2pt}(m=1,2,\cdots,N)$  are not agreement completely, it can be shown by (25) that
$$
\left(\left(U_{\kappa(0)} \otimes C_{o}\right) \left[\begin{array}{c}{\mathbf{0}} \\ {\tilde{x}_{o \Delta}(0)}\end{array}\right]\right)^{T}\left(U_{\kappa(0)} \otimes C_{o}\right) \left[ \begin{array}{c}{\mathbf{0}} \\ {\tilde{x}_{o \Delta}(0)}\end{array}\right]
$$
\begin{equation}\label{eq:26}
\hspace{-60pt}=\tilde{x}_{o \Delta}^{T}(0)\left(I_{N-1} \otimes C_{o}^{T} C_{o}\right) \tilde{x}_{o \Delta}(0)>0.
\end{equation}
By (24) and (26), one has
$$
x_{o}^{T}(0)\left(\left(I_{N}-\frac{1}{N} \mathbf{1 1}^{T}\right) \otimes C_{o}^{T} C_{o}\right) x_{o}(0)>0.
$$
In this case, the given energy budget $J_{\text{e}}^*$  can be denoted by
\begin{equation}\label{eq:27}
J_{\mathrm{e}}^{*}=x_{o}^{T}(0)\left(\left(I_{N}-\frac{1}{N} \mathbf{1 1}^{T}\right) \otimes \zeta C_{o}^{T} C_{o}\right) x_{o}(0),
\end{equation}
where  $\zeta>0$. For the matrix  $I_{N}-N^{-1} \mathbf{1} \mathbf{1}^{T}$, zero eigenvalue is simple and all the other eigenvalues are positive. By (23) and (27), $E_{o}^{T} R_{x} \leq \zeta C_{o}^{T} C_{o}$  can ensure that  $J_{\mathrm{e}} \leq J_{\mathrm{e}}^{*}$. Due to
$$
\hspace{-2cm}x_{o}^{T}(0)\left(\left(I_{N}-\frac{1}{N} \mathbf{1 1}^{T}\right) \otimes C_{o}^{T} C_{o}\right) x_{o}(0)
$$
$$
\hspace{2cm}=y_{o}^{T}(0)\left(\left(I_{N}-\frac{1}{N} \mathbf{1 1}^{T}\right) \otimes I_{h}\right) y_{o}(0),
$$
one can obtain that
$$
y_{o}^{T}(0)\left(\left(I_{N}-N^{-1} \mathbf{1} \mathbf{1}^{T}\right) \otimes I_{h}\right) y_{o}(0) E_{o}^{T} R_{x} \leq J_{\text{e}}^{*} C_{o}^{T} C_{o}.
$$
Based on the above analysis, the conclusion of Theorem 2 can be obtained.$\blacksquare$

\section{Jointly connected communication topology cases}\label{section5}
This section investigates limited-budget output consensus design problems for multiagent systems with jointly connected communication topologies and gives the corresponding output consensus criteria in terms of the Cauchy convergence criterion and Barbalat's lemma.\par

It is assumed that the time interval $[t_m,t_{m+1})$  consists of a series of non-overlapping contiguous subintervals as follows
$$
\begin{aligned}
\left[t_{m}^{0}\hspace{-1pt}, t_{m}^{1}\right)\hspace{-1pt},\left[t_{m}^{1}\hspace{-1pt}, t_{m}^{2}\right)\hspace{-1pt}, \cdots\hspace{-1pt},
\hspace{-2pt}\left[t_{m}^{k_{m}-1}, t_{m}^{k_{m}}\right)\hspace{-2pt}, \hspace{1pt} t_{m}^{0}=t_{m},\hspace{2pt}t_{m}^{k_{m}}=t_{m+1},
\end{aligned}
$$
where $t_{m}^{i+1}\hspace{-2pt}-\hspace{-2pt}t_{m}^{i}\hspace{-2pt}\geq\hspace{-2pt}T_{\mathrm{d}}\left(i=0,1, \cdots\hspace{-1pt}, k_{m}-1\right)$  and $m$  is a nonnegative integer. The communication topologies switch at $t_{m}^{0}, t_{m}^{1}, \cdots, t_{m}^{k_{m}-1}$  and is time-invariant during $\left[t_{m}^{i}, t_{m}^{i+1}\right) \left(i=0,1, \cdots, k_{m}-1\right)$. For jointly connected communication topology cases, the topologies $G_{t_{m}^{0}}, G_{t_{m}^{1}}, \cdots, G_{t_{m}^{k_m-1}}$ can be unconnected, but their union is connected. Let $L_{\kappa\left(t_{m}^{0}\right)}, L_{\kappa\left(t_{m}^{1}\right)}, \cdots, L_{\kappa\left(t_{m}^{k_m-1}\right)}$ be the Laplacian matrices of  $G_{t_{m}^{0}}, G_{t_{m}^{1}}, \cdots, G_{t_{m}^{k_m-1}}$, then $L_{\kappa\left(t_{m}\right)}\hspace{-4pt}=\hspace{-4pt}\sum_{i=0}^{k_{m}-1} L_{\kappa\left(t_{m}^{i}\right)}$  is the Laplacian matrix of a connected topology if the communication topologies during $[t_m,t_{m+1})$  are jointly connected. Here, we set that the transformation  $U_\kappa$ is time-invariant. Let $U_{\kappa}^{T} L_{\kappa\left(t_{m}^i\right)} U_{\kappa}=\operatorname{diag}\left\{\mathbf{0}, \Delta_{\kappa\left(t_{m}^i\right)}\right\}$ with  $\Delta_{\kappa\left(t_{m}^i\right)}=\tilde{U}_{\kappa}^{T} L_{\kappa\left(t_{m}^{i}\right)} \tilde{U}_{\kappa}$, then the following conclusion can be obtain directly.\par

\begin{lem}\label{Lemma 3}
$\Delta_{\kappa\left(t_{m}\right)}=\sum_{i=0}^{k_{m}-1} \Delta_{\kappa\left(t_{m}^i\right)}$  is symmetric and positive definite and its eigenvalues are nonzero eigenvalues of  $L_\kappa(t_m)$.
\end{lem}\par
In the following, a limited-budget output consensus design criterion for jointly connected communication topology cases is proposed.\par
\begin{thm}\label{Theorem 3}
For any given  $J_{\text{e}}^*>0$, multiagent system (1) with jointly connected cummnication topologies is limited-budget output consensualizable by consensus protocol (2) if $(E_o,A_o,C_o)$  is detectable, $(E_o,A_o,B_o)$  is stabilizable, and there exist $\widehat{R}_{x}$ and $\widehat{R}_{z}$ such that
$$
\hspace{-2.8cm}(\mathrm{I}) \operatorname{rank} \left[ \begin{array}{cc}{E_{o}} & {\mathbf{0}} \\ {A_{o}} & {E_{o}}\end{array}\right]=h+\operatorname{rank}\left(E_{o}\right),
$$
$$
\hspace{-0.01cm}(\mathrm{II}) \left\{\begin{array}{l}
\hspace{-0.1cm}{E_{o}^{T} \widehat{R}_{x}=\widehat{R}_{x}^{T} E_{o} \geq 0},\\
\hspace{-0.15cm}{\widehat{R}_{x}^{T} A_{o}+A_{o}^{T} \widehat{R}_{x} \leq 0},\\
\hspace{-0.15cm}\widehat{R}_{x}^{T} A_{o}+A_{o}^{T} \widehat{R}_{x}-C_{o}^{T} C_{o}<0,\\
\hspace{-0.15cm}y_{o}^{T}(0)\left(\left(I_{N}\hspace{-3pt}-\hspace{-3pt}N^{-1} \mathbf{1 1}^{T}\right)\hspace{-3pt}\otimes\hspace{-3pt}I_{h}\right) y_{o}(0) E_{o}^{T} \widehat{R}_{x}\hspace{-3pt}\leq\hspace{-3pt}J_{\text{e}}^{*} C_{o}^{T} C_{o},
\end{array}\right.
 $$
 $$
\hspace{-1.75cm}(\mathrm{III}) \left\{\begin{array}{l}
\hspace{-0.1cm}{\widehat{R}_{z}^{T} E_{o}^{T}=E_{o} \widehat{R}_{z} \geq 0},\\
\hspace{-0.15cm}\widehat{\Theta}_{m}=\left[ \begin{array}{ccc}{\widehat{\Theta}_{11}} & {\widehat{\Theta}_{12}^{m}} & {0.5 B_{o} M} \\ {*} & {\widehat{\Theta}_{22}} & {\mathbf{0}} \\ {*} & {*} & {-M}\end{array}\right]<0,\\
\hspace{-0.15cm}A_{o} \widehat{R}_{z}+\widehat{R}_{z}^{T} A_{o}^{T} \leq 0,
\end{array}\right.
 $$
where $\lambda_{m}=\lambda_{\min }, \lambda_{\max }$  and
$$
\begin{aligned} \widehat{\Theta}_{11} &=A_{o} \hat{R}_{z}+\widehat{R}_{z}^{T} A_{o}^{T}-B_{o} B_{o}^{T}, \\
\widehat{\Theta}_{12}^{m} &=-0.5 \lambda_{m} \widehat{R}_{x}^{-T} C_{o}^{T} C_{o}, \\
\widehat{\Theta}_{22} &=\widehat{R}_{x}^{T} A_{o}+A_{o}^{T} \widehat{R}_{x}-C_{o}^{T} C_{o}. \end{aligned}
$$\par
\end{thm}

In this case, $K_{u}\hspace{-3pt}=\hspace{-3pt}-B_{o}^{T} \widehat{R}_{z}^{-1} / 2$ and  $K_{z}\hspace{-3pt}=\hspace{-3pt}-\lambda_{\min }^{-1} \widehat{R}_{x}^{-T} C_{o}^{T} / 2$.\par

{\it Proof}: Construct the following Lyapunov function candidate
$$
V_{x}(t)=\hat{x}_{o \Delta}^{T}(t)\left(I_{N-1} \otimes E_{o}^{T} \widehat{R}_{x}\right) \hat{x}_{o \Delta}(t),
$$
where $E_{o}^{T} \widehat{R}_{x}=\widehat{R}_{x}^{T} E_{o} \geq 0$.  Let  $K_{z}=-\lambda_{\min }^{-1} \widehat{R}_{x}^{-T} C_{o}^{T} / 2$, then by taking the time derivative of  $V_x(t)$, one can see by (7) that
$$
\dot{V}_{x}(t)=\hat{x}_{o \Delta}^{T}(t)\Big(I_{N-1} \otimes\left(\widehat{R}_{x}^{T} A_{o}+A_{o}^{T} \widehat{R}_{x}\right)
$$
\begin{equation}\label{eq:28}
\hspace{10pt}-\lambda_{\min }^{-1} \Delta_{\kappa(t)} \otimes C_{o}^{T} C_{o} \Big) \hat{x}_{o \Delta}(t).
\end{equation}
Since $\Delta_{\kappa(t)}=\tilde{U}_{\kappa}^{T} L_{\kappa(t)} \tilde{U}_{\kappa}$  and  $L_{\kappa(t)}$ is the Laplacian matrix of an undirected communication topology, $\Delta_{\kappa(t)}$  is symmetric and positive definite and its eigenvalues are the ones of  $L_{\kappa(t)}$  except one zero eigenvalue. Furthermore, since $\widehat{R}_{x}^{T} A_{o}+A_{o}^{T} \widehat{R}_{x} \leq 0$  and $\widehat{R}_{x}^{T} A_{o}+A_{o}^{T} \widehat{R}_{x}-C_{o}^{T} C_{o}<0$,  one can find by (28) that
\begin{equation}\label{eq:29}
\dot{V}_{x}(t) \leq \hat{x}_{o \Delta}^{T}(t)\left(-\lambda_{\min }^{-1} \Delta_{\kappa(t)} \otimes C_{o}^{T} C_{o}\right) \hat{x}_{o \Delta}(t) \leq 0,
\end{equation}
which means that $V_x(t)$  converges to a finite value as time tends to infinity.\par

In the following, it is shown that $V_x(t)$ converges to zero as time tends to infinity.  By the Cauchy convergence criterion, for the infinite sequence $V\left(t_{m}\right)\hspace{2pt}(m=0,1, \cdots)$ and any  $\delta>0$, there exists an integer $M>0$  such that the following inequality holds for  $\forall m>M : V\left(t_{m}\right)-V\left(t_{m+1}\right)<\delta$; that is,
\vskip  0.05cm
$$
-\int_{t_{m}}^{t_{m+1}} \dot{V}(t) \mathrm{d} t<\delta.
$$
Thus, one can obtain that
\begin{equation}\label{eq:30}
-\int_{t_{m}^{0}}^{t_{m}^{1}} \dot{V}(t) \mathrm{d} t-\int_{t_{m}^{1}}^{t_{m}^{2}} \dot{V}(t) \mathrm{d} t-\cdots-\int_{t_{m}^{k_{m}-1}}^{t_{m}^{k_{m}}} \dot{V}(t) \mathrm{d} t<\delta.
\end{equation}
Since $T_{\text{d}}$ is the minimum dwell time, for any  $i \in\left\{0,1, \cdots, k_{m}-1\right\}$, it can be found by (29) that
\begin{equation}\label{eq:31}
-\int_{t_{m}^{i}}^{t_{m}^{i}+T_{\text{d}}} \dot{V}(t) \mathrm{d} t \leq-\int_{t_{m}^{i}}^{t_{m}^{i+1}} \dot{V}(t) \mathrm{d}t.
\end{equation}
From (30) to (31), one has
$$
\sum_{i=0}^{k_{m}-1} \int_{t_{m}^{i}}^{t_{m}^{i}+T_{\text{d}}} \hat{x}_{o \Delta}^{T}(t)\left(\lambda_{\min }^{-1} \Delta_{\kappa\left(t_{m}^{i}\right)} \otimes C_{o}^{T} C_{o}\right) \hat{x}_{o \Delta}(t) \mathrm{d} t<\delta.
$$
Hence, one can show that
\begin{equation}\label{eq:32}
\hspace{-0.1pt}\lim _{t \rightarrow \infty}\hspace{-5pt}\sum_{i=0}^{k_{m}-1}\hspace{-6pt}\int_{t}^{t+T_{\text{d}}}\hspace{-5pt}\hat{x}_{o \Delta}^{T}(t)\hspace{-3pt}\left(\lambda_{\min }^{-1} \Delta_{\kappa\left(t_{m}^{i}\right)}\hspace{-3pt}\otimes\hspace{-2pt}C_{o}^{T} C_{o}\right)\hspace{-2pt}\hat{x}_{o \Delta}\hspace{-2pt}(t)\mathrm{d}t\!=\!0.\hspace{-3.5pt}
\end{equation}
Due to  $\dot{V}(t)\hspace{-5pt}\leq\hspace{-5pt}0$, $\hat{x}_{o\Delta}(t)$ is bounded. By (7),  $\dot{\hat{x}}_{o \Delta}(t)$ is bounded. Hence, one can obtain that $\hat{x}_{o \Delta}^{T}(t)\left(\lambda_{\min }^{-1} \Delta_{\kappa\left(t_{m}\right)}\right.$ $\otimes C_{o}^{T} C_{o}\left)\right. \hat{x}_{o \Delta}(t)$ is uniformly continuous. From Barbalat's lemma in [43], one can find that
\begin{equation}\label{eq:33}
\lim _{t \rightarrow \infty} \hat{x}_{o \Delta}^{T}(t)\left(\lambda_{\min }^{-1} \Delta_{\kappa\left(t_{m}\right)} \otimes C_{o}^{T} C_{o}\right) \hat{x}_{o \Delta}(t)=0.
\end{equation}
By Lemma 3, there exists an orthonormal matrix  $U_{t_{m}}$ such that  $U_{t_{m}}^{T} \Delta_{\kappa\left(t_{m}\right)} U_{t_{m}}=\operatorname{diag}\left\{\lambda_{t_{m}, 2}, \lambda_{t_{m}, 3}, \cdots, \lambda_{t_{m}, N}\right\}>0$. From (32) and (33), one can see that
$$
\begin{aligned}
&\hspace{-15pt}\lim _{t \rightarrow \infty} \lambda_{\min }^{-1} \lambda_{t_{m}, i}\left(U_{t_{m}}^{T} \hat{x}_{o \Delta}^{i}(t)\right)^{T} C_{o}^{T} C_{o}
\\ &\hspace{62pt}\times\left(U_{t_{m}}^{T} \hat{x}_{o \Delta}^{i}(t)\right)=0\;\;(i=2,3, \cdots, N).
\end{aligned}
$$
Because $\left(E_{o}, A_{o}, C_{o}\right)$ is detectable and $U_s$ is nonsingular,  $E_{o}^{T} \widehat{R}_{x}=\widehat{R}_{x}^{T} E_{o} \geq 0$, $\widehat{R}_{x}^{T} A_{o}+A_{o}^{T} \widehat{R}_{x} \leq 0$ and  $\widehat{R}_{x}^{T} A_{o}+A_{o}^{T} \widehat{R}_{x}-C_{o}^{T} C_{o}<0$ in Condition (I) can guarantee that  $\lim _{t \rightarrow \infty} \hat{x}_{o \Delta}(t)=\mathbf{0}$. Similarly, one can show that $\lim _{t \rightarrow \infty} \hat{z}_{o \Delta}(t)=\mathbf{0}$ if Condition (II) holds. Furthermore, the proofs of the regular, impulse-free and limited-budget property are similar to Theorem 2. Thus, the conclusions of Theorem 3 can be obtained.$\blacksquare$\par

\section{Numerical simulation}\label{section7}
In this section, we present two simulation examples to illustrate the validity of the  theoretical results given in the above section and consider a singular system with five agents.
 \vspace{-0.7em}
\begin{figure}[!htb]
\begin{center}
\includegraphics[width=6.5cm]{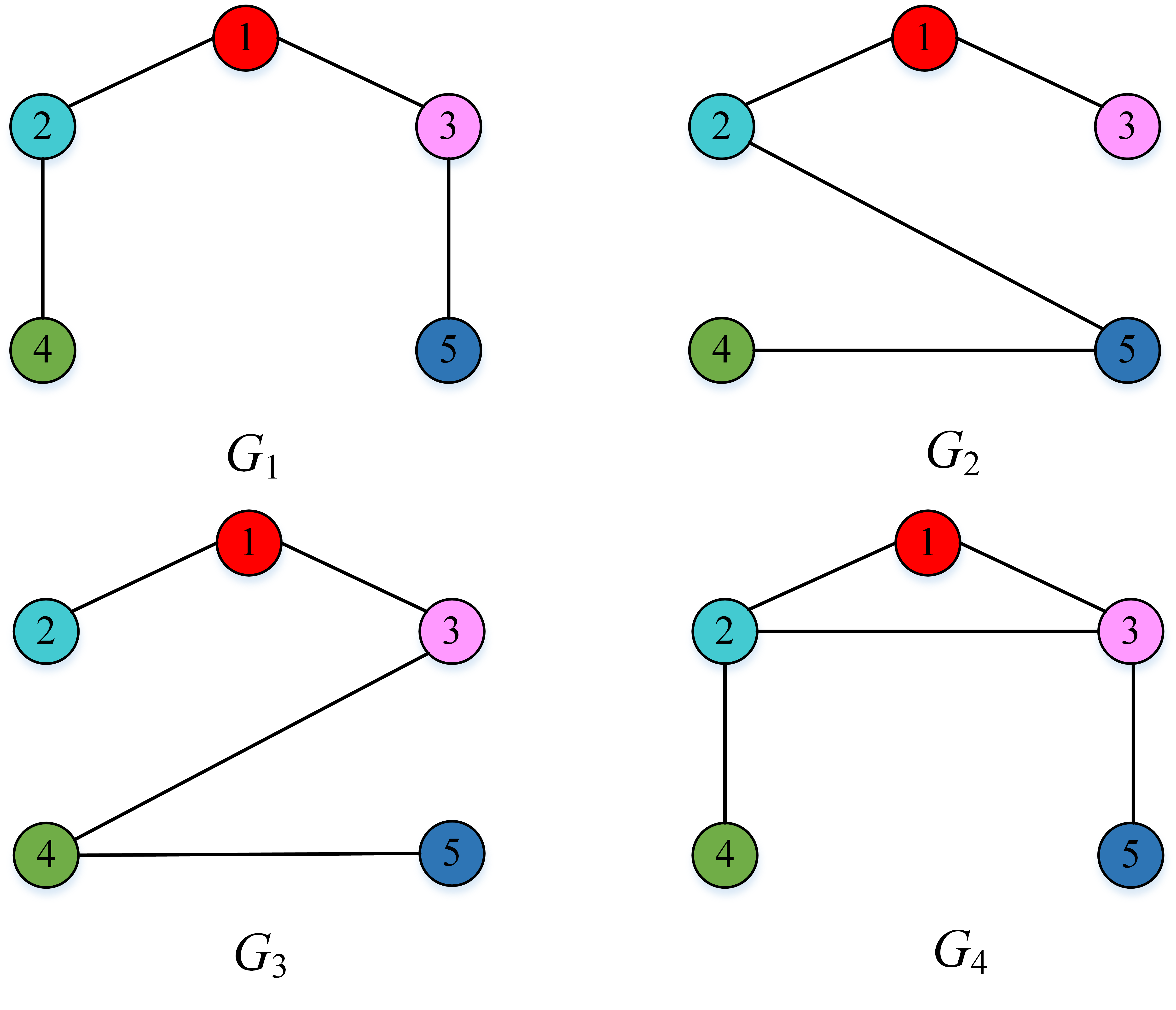}
\caption{Switching connected topologies.}\label{fig1}
\end{center}\vspace{-0.5em}
\end{figure}
\vspace{-1.0em} 
\vskip 0.15cm
 \begin{exm}\label{example1}
({\it Switching connected case}) The system matrices of each agent in multiagent system \eqref{eq:1}  are set as
\[E=\left[ {\begin{array}{*{20}{c}}
   -0.8 &  -2.4  &  2.4  & -2.4 &   1.6 &  0\\
    0.6 &   0.4667  & -0.8  &  0.4667 &   0.3556 &  1\\
    1.2 &  -2.4  & -0.6  & -2.4 &   1.6 &  3\\
    1.2 &  -1.0667  & -1.6  & -0.0667 &   0.7111 &  2\\
    0 &   0  &  0  &  1.5 &   1.0 &  0\\
    2.4 &  -0.1333  & -3.2  & -1.1333 &   0.4222 &  4
\end{array}} \right],\]

\[A\hspace{-1pt}=\hspace{-1pt}\left[{\begin{array}{*{20}{c}}
    3.4  &   4.5333  &   -3.2  &   2.5333  &  -3.0222   &  0\\
    2.4  &   6.8667  &    0.8  &   1.8667  &  -3.2444   & -3\\
    4.2  &   6.6     &   -2.6  &     -3.9  &  -2.4   &  3\\
    2  &   2     &      0  &     -4  &        0   &  2\\
    3  &   7     &    3  &      4  &  -3.6667   &  -6\\
    3.8  &   5.7333  &   -2.4  &  -8.7667  &  -0.4889   &  6\end{array}} \right],
\]
\[B = \left[ {\begin{array}{*{20}{c}}
  -35  &  -12  &  -14  &   -4  &  -33  &    9\\
    18 &    25 &   14  &   12  &   51  &    7
\end{array}} \right]^T,\]
\[
\hspace{0.2cm}C = \left[ {\begin{array}{*{20}{c}}
     -1 & -4.3333  & 3  &  -2.3333 &   2.8889   & 0\\
     -1 & -2.0     & 3  &  -3.5    &   1.3333   &  0
\end{array}} \right],
\]
and the system  can be decomposed into an  observable system by the invertible matrix $U_o$ as
\[U_o=\left[{\begin{array}{*{20}{c}}
       4    &          0         &     3         &     0       &       0       &       0\\
       0    &          4         &     0         &     1       &       2       &       0\\
       3    &          0         &     1         &     3       &       0       &       0\\
       0    &          2         &    0          &     2        &       0      &       0\\
       0    &          9         &     0         &     0       &       3       &       0\\
       0    &          0         &     0         &     4       &       0       &       1\\
\end{array}} \right].
\]

Fig. 1 shows four different undirected connected topologies,  which are chosen as the switching topologies set $\kappa $. For the convenience of analysis, the  communication weights among five agents of each topology are assumed to be 0-1. Fig. 2 indicates that the switching signal is random.

Let
$$M=\left[ \begin{matrix}
   1 & 0  \\
   0 & 1  \\
\end{matrix}\right], $$

\begin{figure}[!htb]
\begin{center}
\includegraphics[width=7.0cm]{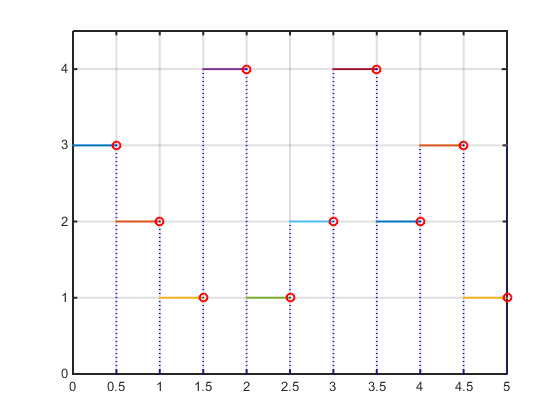}
\put (-192, 70) {\rotatebox{90} {{\scriptsize $\kappa (t)$}}}
\put (-104, 0) {{ \scriptsize {\it t}~/~s}}
\caption{Switching signal for switching connected case.}\label{fig3}
\end{center}\vspace{-1.0em}
\end{figure}
\begin{figure}[!htb]
\begin{center}
\scalebox{0.48}[0.48]{\includegraphics{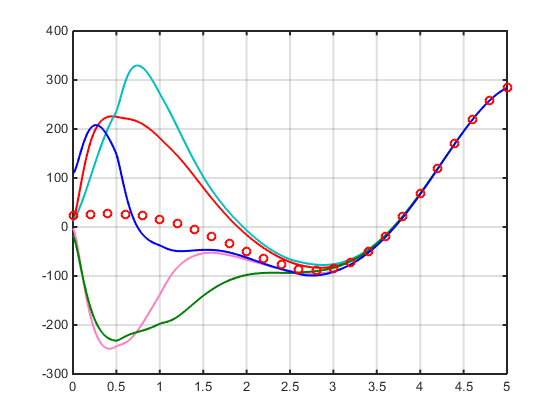}}
\put (-198, 42) {\rotatebox{90} {{\scriptsize $y_{m1}(t)~(m = 1,2, \cdots ,5)$}}}
\put (-107, 0) {{ \scriptsize {\it t}~/~s}}
\end{center}\vspace{-1.40em}
\end{figure}
\begin{figure}[!htb]
\begin{center}
\scalebox{0.48}[0.48]{\includegraphics{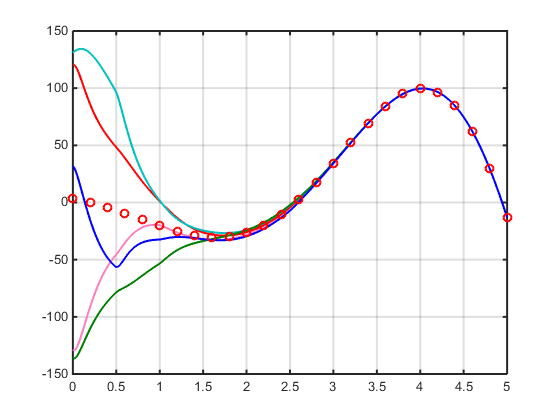}}
\put (-198, 42) {\rotatebox{90} {{\scriptsize $y_{m2}(t)~(m = 1,2, \cdots ,5)$}}}
\put (-107, 0) {{ \scriptsize {\it t}~/~s}}
\caption{Output trajectories for switching connected case.}\label{fig3}
\end{center}\vspace{-1.00em}
\end{figure}
\begin{figure}[!htb]
\begin{center}
\includegraphics[width=7.2cm]{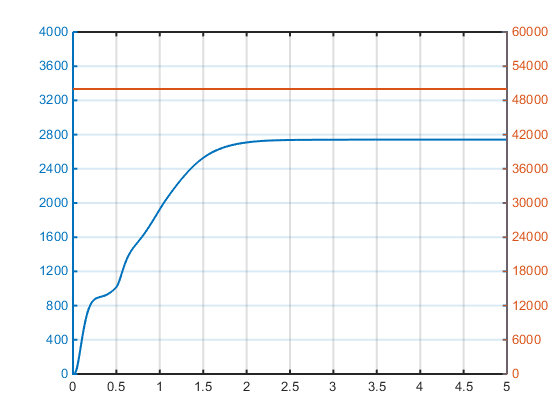}
\put (-90, 125) {{\scriptsize $J_{\text{e}}^{*}$}}
\put (-90, 106) {{\scriptsize ${J_{\text{e}}\left( t \right)}$}}
\put (-105, -5) {{ \scriptsize {\it t}~/~s}}
\caption{Trajectories of ${J_{\text{e}}\left( t \right)}$ and $J_{\text{e}}^{*}$.}\label{fig4}
\end{center}\vspace{-1.0em}
\end{figure}

\hspace{-10pt}and the energy budget is set as $J_\text{e}^{*}=50000$, then  one can obtain from Theorem 2 that
\[{{R}_{x}}=\left[ \begin{matrix}
   20.8005  &    -2.6807   &         0\\
   -2.6807  &   14.8810    &       0\\
  -18.1199  &  -12.2003    &  -10.0022
\end{matrix} \right],\]

\[{{R}_{z}}=\left[ \begin{matrix}
    1.0833  &  1.5299    &     0\\
    1.5299  &  2.2577    &     0\\
 -909.7727  &-910.3096   &-911.3356
\end{matrix} \right],\]
\[{{K}_{u}}= \left[ \begin{matrix}
     7.3118  & -5.1748 &  -0.0027\\
   11.0466   & -8.1502  & 0.0011
\end{matrix} \right],\]

\[{{K}_{z}}= \left[ \begin{matrix}
   -0.3686  & -0.4183  & 0\\
   -0.2873  &  0.2121  & 0
\end{matrix} \right]^{T}.\]

We set the initial state value of each agent as follows:
\[\begin{array}{l}
 {{x}_{1}}(0)={{[60, -26, 81, -6, -102, 38]}^{T}},\text{ } \\
 {{x}_{2}}(0)={{[64, -81,  75, -16, -186, 24]}^{T}}, \\
  {{x}_{3}}(0)={{[-60, 103,-6, 62, 189, 37]}^{T}}, \\
  {{x}_{4}}(0)={{[ -68, 49, -30, 48, 114, 19]}^{T}}, \\
 {{x}_{5}}(0)={{[52, 38, 87, 54, 66, 52]}^{T}}.\text{  } \\
\end{array}\]

In Fig. 3, the output trajectories of the descriptor multiagent system are depicted, where the red circle markers describe the trajectory of the output consensus function in Theorem 4. Fig. 4 shows the trajectory of the  energy cost function $J_\text{e}(t)$ with $\hbar$ = 5. One can see that output trajectories of all agents converge to the curve formed by circle markers and $J_\text{e}(t)<J_\text{e}^*$, which means that the descriptor multiagent system  achieves limited-budget output consensus.
\end{exm}

\vskip -0.250cm
\begin{exm}\label{example2}
({\it Jointly connected case}) In this case, the system matrices of each agent in multiagent system \eqref{eq:1}  are set as
 \[E=\left[ {\begin{array}{*{20}{c}}
   -8  &   9  &   6  &  12    &   -9   &  -17\\
   -5  &   9  &   4  &  11    &   -6   &  -14\\
  -12  &  14.5  &   9  &  21.25 &  -12   &  -28\\
  -10  &  14  &   8  &  22    &  -12   &  -28\\
    0  &   3  &   0  &   4    &    1   &   -4\\
  -10  &  14  &   8  &  21    &  -12   &  -27
\end{array}} \right],\]
\vskip -0.250cm
\[A=\left[\hspace{-3pt}{\begin{array}{*{20}{c}}
   13 &  -11.5 &   -8  & -15.75 &   11 &   22\\
   -4 &   1.5  &   4   &  1.75  &  -1  &  -2\\
   31 &  -38.5 &  -21 &  -55.75 &  33  &  75\\
  -20 &   27 &   16  &  43.5 &  -22  & -56\\
   26 &  -41 &  -19  & -64 &  34  &  84\\
  -24 &   31.5 &   19 &   50.25 &  -26  & -65
\end{array}} \right],
\]
\[B = \left[ {\begin{array}{*{20}{c}}
     -11  & -11  &  10  &  -6  &  17  & -11\\
     30   & -3   & 58   & -2   & 15   & -5
\end{array}} \right]^T,\]
\[
\hspace{0.2cm}C = \left[ {\begin{array}{*{20}{c}}
   -32 &  36    & 24   &   58   & -32  &-76\\
   -20 & 22.5  & 15   &29.75  &  -20  &-41
 \end{array}} \right],\]
and the system  can be decomposed into an  observable system by the invertible matrix $U_o$ as
\[U_o=\left[{\begin{array}{*{20}{c}}
       1      &        0       &       3        &      0        &      0       &       1\\
       0      &        3       &       0        &      1        &      2       &       0\\
       3      &        0       &       4        &      0        &      1       &       0\\
       0      &        2       &       0        &      2        &      0       &       0\\
       1      &        0       &       0        &      0        &      3       &      -1\\
       0      &        3       &       0        &      2        &      0       &       0
 \end{array}} \right].
\]

\begin{figure}[!htb]
\begin{center}
\includegraphics[width=6.5cm]{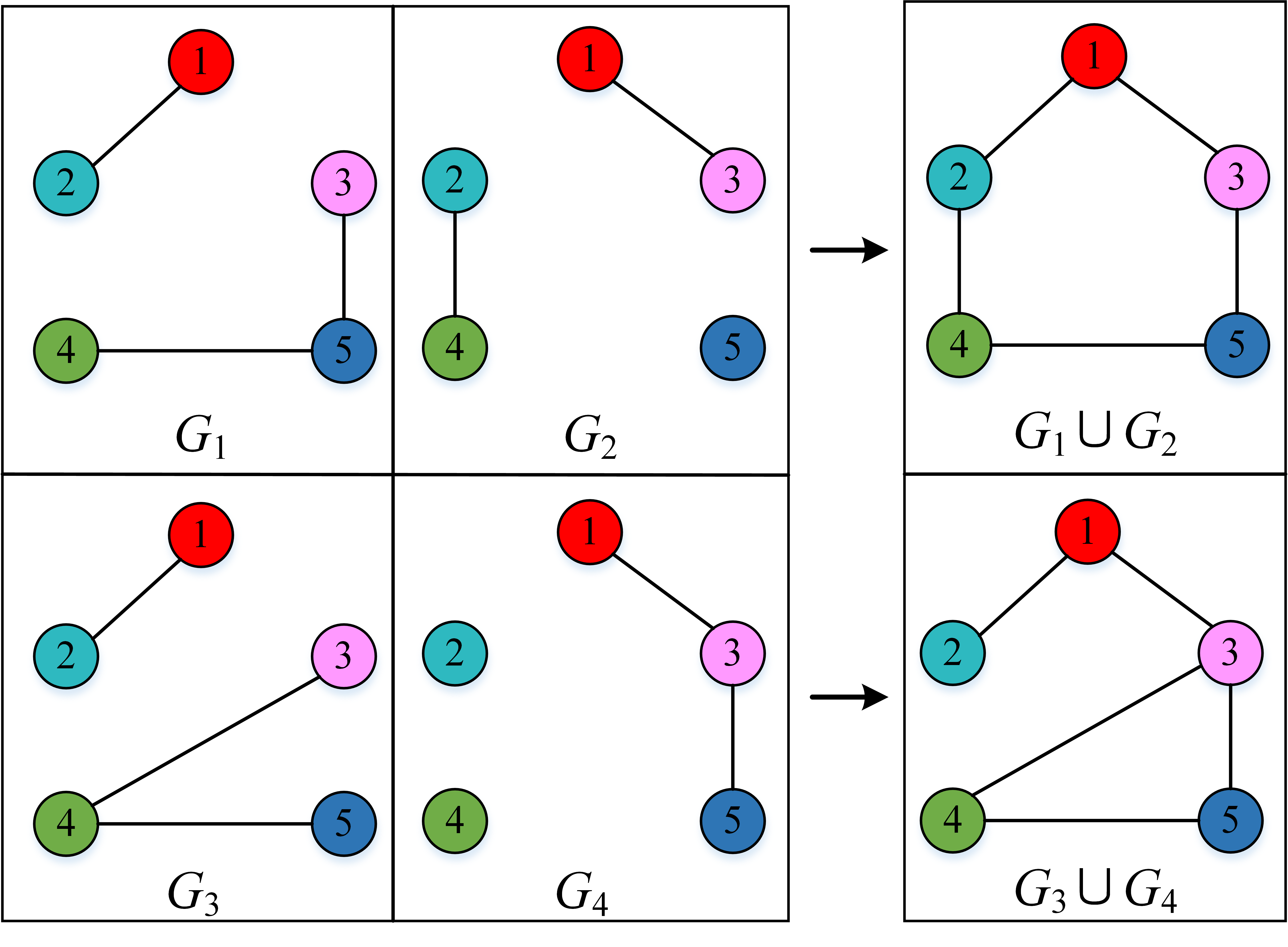}

\caption{Jointly connected topologies.}\label{fig5}
\end{center}\vspace{-0.5em}
\end{figure}

Fig. 5 shows two different jointly connected topologies and each topology is composed of two unconnected topologies. For the convenience of analysis, the  communication weights among these agents of each topology are assumed to be 0-1.  The switching order of four topologies is designed as ${G_1} \to {G_2} \to {G_3} \to {G_4} \to {G_1} \to {G_2} \cdots $; that is, the switching process is repeated every four times as shown in Fig. 6.

Let
$$M=\left[ \begin{matrix}
   1 & 0  \\
   0 & 1  \\
\end{matrix} \right],$$ and the energy budget is given as $J_\text{e}^{*}=10000$.
Then, by Theorem 3, one can show that
\[{\widehat{R}_{x}}=\left[ \begin{matrix}
  26.9957    &     0    &     0\\
         0  & 26.9957   &     0\\
         0   &      0   & -6.7489
\end{matrix} \right],\]

\[{\widehat{R}_{z}}=\left[ \begin{matrix}
   56.8087  &       0   &      0\\
         0  & 56.8087   &      0\\
         0  &       0   &-14.2022
\end{matrix} \right],\]
\[{{K}_{u}}= \left[ \begin{matrix}
     -0.1320 &   0.0440 &  -0.3169\\
     -0.1056 &   0.0264 &   0.1760

\end{matrix} \right],\]

\[{{K}_{z}}= \left[ \begin{matrix}
     0.2856  & -0.1428   &      0\\
   -0.1785   & -0.1428   &      0\\
\end{matrix} \right]^{T}.\]

We set the initial state value of each agent as follows:
\[\begin{array}{l}
 {{x}_{1}}(0)={{[ 2,    -7,   -9,    -6,    6,   -14]}^{T}},\text{ } \\
 {{x}_{2}}(0)={{[ 14,    41,   15,    42,    -4,    52]}^{T}}, \\
  {{x}_{3}}(0)={{[ 11,   -29,  18,   -18,    15,   -34]}^{T}}, \\
  {{x}_{4}}(0)={{[ 2,    19,   -18,     2,    -7,     8]}^{T}}, \\
 {{x}_{5}}(0)={{[ 16,    -3,   33,   -14,    27,   -21]}^{T}}.\text{  } \\
\end{array}\]
\vspace{-1.5em}
\begin{figure}[!htb]
\begin{center}
\includegraphics[width=7.0cm]{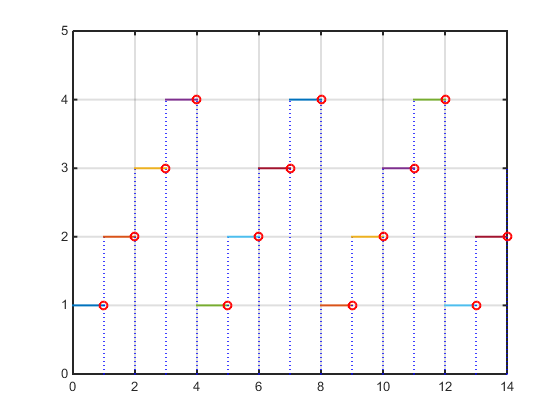}
\put (-192, 70) {\rotatebox{90} {{\scriptsize $\kappa (t)$}}}
\put (-104, 0) {{ \scriptsize {\it t}~/~s}}
\caption{Switching signal for jointly connected cases.}\label{fig6}
\end{center}\vspace{-1.1em}
\end{figure}
\begin{figure}[!htb]
\begin{center}
\scalebox{0.48}[0.47]{\includegraphics{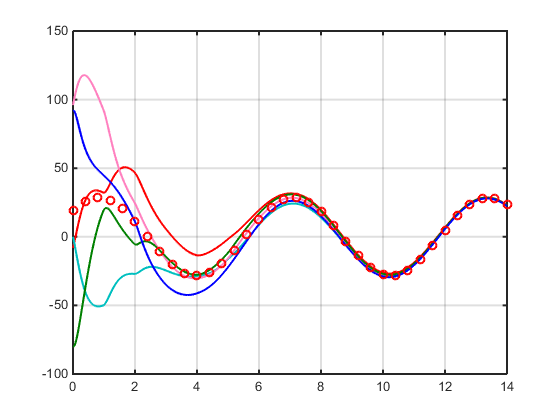}}
\put (-198, 38) {\rotatebox{90} {{\scriptsize $y_{m1}(t)~(m = 1,2, \cdots ,5)$}}}
\put (-105, 0) {{ \scriptsize {\it t}~/~s}}
\end{center}\vspace{-2.10em}
\end{figure}
\begin{figure}[!htb]
\begin{center}
\scalebox{0.48}[0.47]{\includegraphics{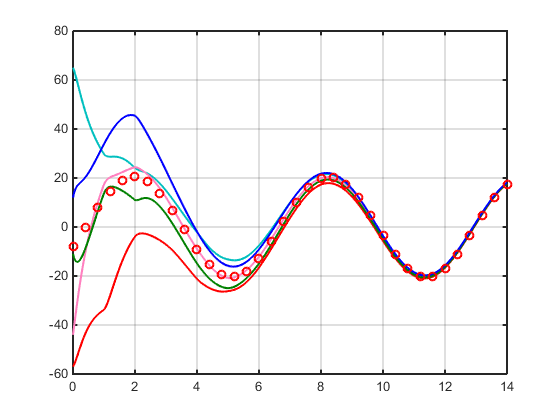}}
\put (-198, 38) {\rotatebox{90} {{\scriptsize $y_{m2}(t)~(m = 1,2, \cdots ,5)$}}}
\put (-105,0) {{ \scriptsize {\it t}~/~s}}
\caption{Output trajectories for jointly connected case.}\label{fig7}
\end{center}\vspace{-0.10em}
\end{figure}

%
Fig. 7 depicts the output trajectories of this descriptor multiagent system, where the red circle markers depict the curves of the output consensus function shown in Theorem 4. Fig. 8 shows the trajectory of the  energy cost function $J_\text{e}(t)$ with $\hbar$ = 14. From these figures, two output trajectories of all agents  converge to the curves formed by the red circle markers and $J_\text{e}(t)<J_\text{e}^*$, which means that this descriptor multiagent system  achieves limited-budget output consensus.

\begin{figure}[!htb]
\begin{center}
\includegraphics[width=7.2cm]{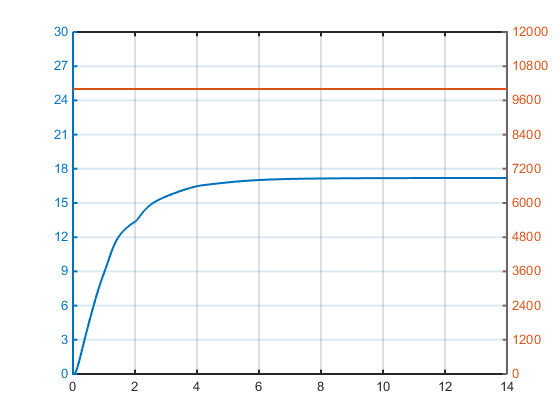}
\put (-90, 127) {{\scriptsize $J_{\text{e}}^{*}$}}
\put (-90, 93) {{\scriptsize ${J_{\text{e}}\left( t \right)}$}}
\put (-107, 1) {{ \scriptsize {\it t}~/~s}}
\caption{Trajectories of ${J_{\text{e}}\left( t \right)}$ and $J_{\text{e}}^{*}$.}\label{fig8}
\end{center}\vspace{-0.5em}
\end{figure}

\end{exm}


\section{Conclusions}\label{section6}
For descriptor multiagent systems with both switching connected communication topologies and jointly connected communication topologies, a descriptor dynamic output feedback consensus protocol with an energy constraint was proposed to realize limited-budget output consensus. By the matrix inequality tool, sufficient conditions for limited-budget output consensus design of multiagent systems with switching connected communication topologies were presented, where a new two-step design approach was given to deal with the nonlinear coupled design problems of two gain matrices and those sufficient conditions are checkable since they are independent of the number of agents. Furthermore, by combining the Cauchy convergence criterion and Barbalat's lemma, limited-budget output consensus design criteria for jointly connected communication topology cases were proposed, where it was required that each agent is Lyapunov stable.  \par

\ifCLASSOPTIONcaptionsoff
\newpage
\fi


\end{document}